\documentclass[english,10pt,aps,pra,twocolumn,groupedaddress,floatfix]{revtex4-1}
\pdfoutput=1
\usepackage[utf8]{inputenc}
\usepackage[T1]{fontenc}
\usepackage{amsmath}
\usepackage{amssymb}
\usepackage{amsfonts}
\usepackage{bm}
\usepackage{bbm}
\usepackage{epsfig}
\usepackage{grffile}

\usepackage{scalefnt}
\usepackage{numprint}

\usepackage[usenames,dvipsnames]{color}
\definecolor{dblue}{rgb}{0,0.1,.6}

\usepackage[colorlinks=true,citecolor=dblue,linkcolor=dblue,urlcolor=dblue]{hyperref}
\usepackage[all]{hypcap}
\newcommand{\Footnote}[1]{\footnote{\unexpanded{#1}}}

\newcommand{\id}{\mathbbm{1}}
\newcommand{\bra}{\langle}
\newcommand{\ket}{\rangle}
\newcommand{\bbra}{\langle\!\langle}
\newcommand{\kket}{\rangle\!\rangle}
\newcommand{\Tr}{\operatorname{Tr}}
\newcommand{\mc}[1]{\mathcal{#1}}
\newcommand{\ud}{\mathrm{d}}
\newcommand{\ue}{\mathrm{e}}
\newcommand{\mri}{\mathrm{i}}

\renewcommand{\O}{\mc{O}}
\newcommand{\E}{\mc{E}}
\newcommand{\M}{\mc{M}}
\renewcommand{\S}{\mc{S}}
\newcommand{\N}{\mc{N}}

\newcommand{\hH}{\hat{H}}
\newcommand{\hh}{\hat{h}}
\newcommand{\hA}{\hat{A}}
\newcommand{\hB}{\hat{B}}
\newcommand{\hg}{\hat{g}}
\newcommand{\hl}{\hat{\ell}}
\newcommand{\hr}{\hat{r}}
\newcommand{\hL}{\hat{L}}
\newcommand{\hLambda}{\hat{\Lambda}}
\newcommand{\hU}{\hat{U}}
\newcommand{\tU}{\tilde{U}}
\newcommand{\hV}{\hat{V}}
\newcommand{\hR}{\hat{R}}
\newcommand{\hsigma}{\hat{\sigma}}
\newcommand{\hX}{\hat{X}}
\newcommand{\hY}{\hat{Y}}
\newcommand{\dm}{{\hat{\rho}}}

\newcommand{\CC}{\mathbb{C}}
\newcommand{\groupU} {\operatorname{U}}
\newcommand{\doub}{{(2)}}
\newcommand{\Avg} {\operatorname{Avg}}
\newcommand{\Var} {\operatorname{Var}}

\newcommand{\mps}{\text{mps}}
\newcommand{\bin}{\text{bi}}
\newcommand{\tL}{\text{L}}
\newcommand{\tR}{\text{R}}

\newcommand{\duke} {Department of Physics, Duke University, Durham, North Carolina 27708, USA}
\newcommand{\dqc}  {Duke Quantum Center, Duke University, Durham, North Carolina 27701, USA}

\newcommand{\Title} {Isometric tensor network optimization for extensive Hamiltonians is free of barren plateaus}
\newcommand{\Authors}
{
\author{Qiang Miao}
\affiliation{\duke}
\affiliation{\dqc}
\author{Thomas Barthel}
\affiliation{\duke}
\affiliation{\dqc}
}
\newcommand{\Date} {February 5, 2024}

\begin{document}

\title{\texorpdfstring{\scalefont{0.95}\Title}{\Title}}
\Authors

\date{\Date}

\begin{abstract}
We explain why and numerically confirm that there are no barren plateaus in the energy optimization of isometric tensor network states (TNS) for extensive Hamiltonians with finite-range interactions which are, for example, typical in condensed matter physics. Specifically, we consider matrix product states (MPS) with open boundary conditions, tree tensor network states (TTNS), and the multiscale entanglement renormalization ansatz (MERA). MERA are isometric by construction and, for the MPS and TTNS, the tensor network gauge freedom allows us to choose all tensors as partial isometries.
The variance of the energy gradient, evaluated by taking the Haar average over the TNS tensors, has a leading system-size independent term and decreases according to a power law in the bond dimension. For a hierarchical TNS (TTNS and MERA) with branching ratio $b$, the variance of the gradient with respect to a tensor in layer $\tau$ scales as $(b\eta)^\tau$, where $\eta$ is the second largest eigenvalue of a Haar-average doubled layer-transition channel and decreases algebraically with increasing bond dimension. The absence of barren plateaus substantiates that isometric TNS are a promising route for an efficient quantum-computation-based investigation of strongly-correlated quantum matter. The observed scaling properties of the gradient amplitudes bear implications for efficient TNS initialization procedures. 
\end{abstract}

\maketitle

\section{Introduction}\label{sec:intro}\vspace{-0.5em}
Rapid recent advances in quantum technology have opened new routes for the solution of hard quantum groundstate problems. Well-controlled quantum devices are a natural platform for the investigation of complex quantum matter \cite{Feynman1982-21}. An important approach are variational quantum algorithms (VQA) \cite{Cerezo2021-3}, in which classical optimization is performed on parametrized quantum circuits. Numerous studies have successfully applied VQA to few-body systems, but applications of generic unstructured or overly expressive VQA to many-body systems face multiple challenges: (a) Current quantum hardware is limited in the available number of qubits and gate fidelities. (b) As with other high-dimensional nonlinear optimization problems, we are typically confronted with complex cost-function landscapes and it is difficult to avoid local minima \cite{Kiani2020_01,Bittel2021-127,Anschuetz2021_09,Anschuetz2022-13}. (c) In contrast to classical computers, the quantum algorithms produce a sequence of probabilistic measurement results for gradients, requiring a large number of shots to achieve precise estimates. For a generic unstructured VQA, gradient amplitudes decay exponentially in the size of the simulated system \cite{McClean2018-9,Cerezo2021-12}. This is the so-called barren plateau problem, which is also associated with cost-function concentration \cite{Arrasmith2022-7,Miao2024_02}. In this case, VQA are not trainable as the inability to precisely estimate exponentially small gradients will result in random walks on the flat landscape.
The barren plateau problem can be resolved if an initial guess close to the optimum and specific optimization strategies are available \cite{Grant2019-3,Zhang2022_03,Mele2022-106,Kulshrestha2022_04,Dborin2022-7,Skolik2021-3,Slattery2022-4,Haug2021_04,Sack2022-3,Rad2022_03,Tao2022_05}, but such resolutions are not universal \cite{Campos2021-103}.

Unstructured circuits like the hardware efficient ansatz and brickwall circuits must be deep in order to cover relevant parts of the Hilbert space \cite{Dankert2009-80,Brandao2016-346,Harrow2023-05}. The high expressiveness of such circuits \cite{Sim2019-2,Nakaji2021-5,Du2022-128} can be seen as the source for the barren plateaus \cite{Holmes2022-3}. The latter can be motivated by typicality properties \cite{Goldstein2006-96,Popescu2006-2} of such states \cite{Ortiz2021-2,Patti2021-3,Sharma2022-128} which can, for example, display volume-law entanglement entropies.
So, it is generally preferable to work with more structured and less entangled classes of states that are adapted to the particular optimization problem in order to balance expressiveness and trainability.
\begin{figure*}[t]
	\label{fig:isoTNS}
	\includegraphics[width=\textwidth]{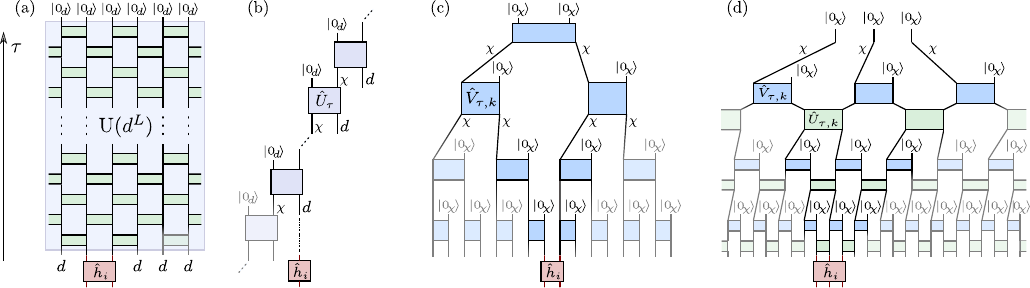}
	\caption{\textbf{Isometric quantum circuits.}
	(a) A generic brickwall quantum circuit for $L$ qudits consisting of layers of nearest-neighbor two-qudit gates.
	(b) MPS quantum circuit with single-site Hilbert space dimension $d$ and bond dimension $\chi$. Using the TNS gauge freedom to bring the MPS into left-canonical form \cite{Schollwoeck2011-326,Barthel2022-112}, the expectation value $\bra\Psi|\hh_i|\Psi\ket$ for an operator acting on site $i$ only depends on the tensors for sites $\tau\geq i$.
	(c,d) Binary 1D TTNS and MERA quantum circuits with branching ratio $b=2$ and bond dimension $\chi$. Only the non-shaded tensors can influence the expectation values of the local operator $\hh_i$ and constitute its \emph{causal cone}.}
\end{figure*}

In this work, we demonstrate that VQA barren plateaus can be avoided for quantum many-body groundstate problems by employing matrix product states (MPS) with open boundary conditions \cite{Baxter1968-9,Accardi1981,Fannes1992-144,White1992-11,Rommer1997,PerezGarcia2007-7,Schollwoeck2011-326}, tree tensor network states (TTNS) \cite{Fannes1992-66,Otsuka1996-53,Shi2006-74,Murg2010-82,Tagliacozzo2009-80}, or the multiscale entanglement renormalization ansatz (MERA) \cite{Vidal-2005-12,Vidal2006,Barthel2010-105}. We refer to these tensor network states (TNS) as being \emph{isometric}, because, all tensors in the network are either isometries by definition (MERA) or one can use the TNS gauge freedom \cite{Barthel2022-112,Evenbly2022-8} to make all tensors isometric (MPS and TTNS). The entanglement structures of these TNS are well-adapted to those in many-body ground states, and classical simulations have established TNS algorithms as valuable tools for the investigation of strongly-correlated quantum matter. However, the classical TNS computation costs grow quickly with increasing TNS bond dimension $\chi$ which controls the achievable approximation accuracy. Especially for quasi-2D and 3D systems, the rapid growth of classical tensor contraction costs in $\chi$ \cite{Miao2021_08} limits investigations of important phenomena such as high-temperature superconductivity and topological order that could be employed for error-protected quantum computation.

Fortunately, variations of isometric TNS can be implemented on quantum computers \cite{Liu2019-1,Miao2021_08,FossFeig2021-3,Niu2022-3,Slattery2021_08}, which may allow us to substantially reduce computation costs in comparison to classical simulations \cite{Miao2021_08,Miao2023_03}. Recently, barren plateaus have been studied using graph techniques for MPS \cite{Liu2022-129,Garcia2023-2023} and ZX-calculus for TNS with bond dimension $\chi=2$ \cite{Zhao2021-5,Martin2023-7} in the energy optimization for (tensor products of) single-site Hamiltonians. Of course, the latter are actually solved by (products of) the single-site ground states and, hence, of no practical relevance \Footnote{For a system of $L$ sites, the expectation value $\bra\Psi|\hh_1\otimes\hh_2\otimes\dotsb\otimes\hh_L|\Psi\ket$ for normalized states $\Psi$ is simply minimized by the tensor product $|\Psi\ket=|\psi_1\ket\otimes\dotsb\otimes|\psi_L\ket$ of the lowest-eigenvalue eigenstates $\psi_i$ of the single-site Hamiltonians $\hat{h}_i$. The same holds for sums $\bra\Psi|\sum_{i=1}^L\hh_i|\Psi\ket$ of single-site Hamiltonians}.

Here, we address the actual (QMA-complete \cite{Kempe2006-35,Oliveira2008-8,Aharonov2009-287,Gottesman2009_05,Bausch2017-18}) groundstate problem for extensive translation-invariant Hamiltonians
\begin{equation}\label{eq:H}\textstyle
	\hH = \sum_i\hh_i\quad\text{with}\quad\Tr\hh_i=0
\end{equation}
encountered in quantum many-body physics, where the finite-range interaction term $\hh_i$ acts non-trivially in the vicinity of site $i$. We explain why the corresponding VQA, minimizing the energy expectation value $\bra\Psi|\hH|\Psi\ket$ with respect to an isometric TNS $|\Psi\ket$ does \emph{not} encounter barren plateaus. Detailed proofs of the analytical results are given in the companion paper \cite{Barthel2023_03}, and they are confirmed here numerically.

The key ideas are the following: Due to the isometric properties of the tensors, the TNS expectation value for a local interaction term $\hh_i$ only depends on tensors in the causal cone of $\hh_i$. The expectation values can be evaluated by propagating causal-cone density operators $\dm$ in the preparation direction (decreasing $\tau$ in Fig.~\ref{fig:isoTNS}) with transition maps $\M$ and/or interaction terms $\hh$ in the renormalization direction (increasing $\tau$) with $\M^\dag$.
To evaluate the variance of the energy gradient for TNS tensors sampled according to the Haar measure, doubled transition quantum channels $\E^\doub:=\Avg\M\otimes\M$ are applied multiple times to $\dm\otimes\dm$ and their adjoints $\E^{(2)\dag}$ to $\hh\otimes\hh$. While the image of $\dm\otimes\dm$ will quickly converge to a unique steady state, we find that the leading contribution from $\hh\otimes\hh$ has a decay factor $(b\eta)^\tau$, where $\eta$ is the second largest eigenvalue of $\E^\doub$ and $b$ the branching ratio of the TNS ($b=1$ for MPS and $b>1$ for TTNS and MERA).

This leads to three key observations: for isometric TNS and extensive Hamiltonians with finite-range interactions, (i) the gradient variance is independent of the total system size rather than exponentially small, (ii) the gradient variance for a tensor in layer $\tau$ of hierarchical TNS decays exponentially in the layer index $\tau$, (iii) the gradient variances decrease according to power laws in the TNS bond dimension $\chi$. Instead of Euclidean gradients in parametrized quantum circuits, we employ Riemannian gradients which greatly simplifies the proofs \cite{Barthel2023_03}.

\section{Riemannian TNS gradients}
All tensors in the considered isometric TNS are either unitaries $\hU$ or partial isometries $\hV:\CC^{N_1}\to\CC^{N_1}\otimes \CC^{N_2}$ with $\hV^\dag\hV=\id_{N_1}$: MERA are isometric by definition \cite{Vidal-2005-12}; the MPS and TTNS can always be brought into isometric form by using the TNS gauge freedom \cite{Schollwoeck2011-326,Barthel2022-112,Evenbly2022-8}.
The isometries $\hV$ can be implemented as partially projected unitaries in the form 
\begin{equation}\label{eq:iso-to-unitary}
	\hV=\hU\,(\id_{N_1}\otimes|0_{N_2}\ket)\quad\text{with}\quad
	\hU\in \groupU(N_1 N_2)
\end{equation}
and an arbitrary reference state $|0_{N_2}\ket\in \CC^{N_2}$.
The TNS energy expectation values can be written in the form 
\begin{equation}\label{eq:energy}
	E(\hU) = \bra\Psi(\hU)|\hH|\Psi(\hU) \ket = \Tr(\hX \tU^\dag \hY \tU ),
\end{equation}
where we explicitly denote the dependence on one of the TNS unitaries $\hU\in\groupU(N)$ and $\tU:=\hU\otimes\id_M$. The Hermitian operators $\hX$ and $\hY$ depend on the other TNS tensors and $\hY$ also comprises the Hamiltonian. The Riemannian energy gradient is then given by \cite{Hauru2021-10,Luchnikov2021-23,Miao2021_08,Wiersema2023-107,Barthel2023_03}
\begin{equation}\label{eq:gradRiemann}
	\hg(\hU) = \partial_{\hU}\bra\Psi|\hH|\Psi\ket =  \Tr_M(\hY\tU\hX - \tU\hX\tU^\dag\hY\tU).
\end{equation}
Averaged according to the Haar measure, $\hg$ vanishes,
\begin{equation*}\textstyle
	\Avg_{\hU}\hg:=\int \ud U\, \hg(\hU) = \frac{1}{2} \int \ud U [\hg(\hU) + \hg(-\hU)] = 0.
\end{equation*}

In order to assess the question of barren plateaus, the Haar variance of the Riemannian gradient \eqref{eq:gradRiemann} can be quantified by
\begin{eqnarray}\label{eq:var}\textstyle 
 	\Var_{\hU}\hg  :=  \Avg_{\hU} \frac{1}{N} \Tr(\hg^\dag\hg).
\end{eqnarray}
We can expand $\hg$ in an orthonormal basis of $N^2$ Hermitian and unitary operators $\{\hsigma_n\}$ with $\hsigma_n^2=\id_N$ such that $\hg=\mri \hU\sum_{n=1}^{N^2}\alpha_n \hsigma_n/N$. On a quantum computer, the rotation-angle derivatives can be determined as energy differences \cite{Miao2021_08,Wiersema2023-107}
\begin{equation}
	\alpha_n=E(\hU\ue^{{\mri\pi\hsigma_n}/{4}}) - E(\hU\ue^{{-\mri\pi\hsigma_n}/{4}}).
\end{equation}
Equation~\eqref{eq:var} then agrees with the rotation-angle variance $\int\ud U\,\frac{1}{N^2}\sum_n\alpha_n^2$, motivating the employed factor $1/N$.

We focus on extensive Hamiltonians \eqref{eq:H} with finite-range interactions $\hh_i$. Let $\tau$ identify one unitary tensor $\hU_\tau$ in the TNS and $\S_\tau$ the set of physical sites $i$ with $\hU_\tau$ in the causal cone (cf.\ Fig.~\ref{fig:isoTNS}). The gradient \eqref{eq:gradRiemann} then takes the form
\begin{equation}
	\hg(\hU_\tau)=\sum_{i\in\S_\tau}\hg^{(i)}_\tau\quad\text{with}\quad
	\hg^{(i)}_\tau := \partial_{\hU_\tau}\bra\Psi|\hh_i|\Psi\ket.
\end{equation}
Averaging over all unitaries of the TNS, the Haar variance of $\hg(\hU_\tau)$ reads
\begin{equation}\label{eq:var0}
	\Var\hg(\hU_\tau) = \sum_{i_1,i_2\in\S_\tau} \Avg\, \frac{1}{N} \Tr\big(\hg_\tau^{(i_1)\dag} \hg_\tau^{(i_2)}\big),
\end{equation}
where $\Avg$ denotes the Haar-average over all remaining TNS tensors besides $\hU_\tau$.

\section{Matrix product states}
Consider MPS of bond dimension $\chi$ for a system of $L$ sites and single-site Hilbert-space dimension $d$,
\begin{equation}
	|\Psi\ket=\!\!\sum_{s_1,\dotsc,s_L=1}^d\!\! \bra 0| \hV_1^{s_1}\hV_2^{s_2}\dotsb \hV_L^{s_L} |0\ket\, |s_1,s_2,\dotsc,s_L\ket.
\end{equation}
Using its gauge freedom, the MPS can be brought to left-canonical (a.k.a.\ left-orthonormal) form \cite{Schollwoeck2011-326,Barthel2022-112}, where the tensors $\hV_\tau$ with $\bra a,s_\tau|\hV_\tau|b\ket:=\bra a|\hV_\tau^{s_\tau}|b\ket$, $s_\tau=1,\dotsc,d$, and $a,b=1,\dotsc,\chi$ are isometries in the sense that $\hV_\tau^\dag \hV_\tau^{} = \id_\chi$. We use Eq.~\eqref{eq:iso-to-unitary} to express them in terms of unitaries with $\hV_\tau=:\hU_\tau\,(\id_\chi\otimes|0_d\ket)$ in the bulk of the system such that $\hU_\tau\in\groupU(\chi\, d)$ \cite{Barthel2023_03}.

For simplicity, let us first address Hamiltonians \eqref{eq:H} with single-site terms $\hh_i=\id_d^{\otimes(i-1)}\otimes\hh\otimes\id_d^{\otimes(L-i)}$. Due to the left-orthonormality, the local expectation value $\bra\Psi|\hh_i|\Psi\ket$ is independent of all tensors $\hU_\tau$ with $\tau<i$ such that $\S_\tau=\{1,\dotsc,\tau\}$. As we chose $\Tr\hh=0$ without loss of generality, all off-diagonal contributions with $i_1\neq i_2$ in Eq.~\eqref{eq:var0} vanish. It remains to evaluate the diagonal contributions with $i_1=i_2=i\leq \tau$: The expectation value has the form \eqref{eq:energy}. In particular,
\begin{subequations}\label{eq:mps-hi-expect}
\begin{align}
	\bra\Psi|\hh_i&|\Psi\ket = \Tr(\hX_\tau^{(i)} \hU_\tau^\dag \hY_\tau^{(i)} \hU_\tau )\quad\text{with}\\
	\hX_\tau^{(i)}&=\M_{\tau+1}\circ\dotsb\circ\M_{L}(|0\ket\bra 0|)\otimes|0_d\ket\bra 0_d|\ \ \text{and}\\
	\hY_\tau^{(i)}&=\M^\dag_{\tau-1}\circ\dotsb\circ\M^\dag_{i+1}(\hL^{(i)})\otimes\id_d,\quad
\end{align}
\end{subequations}
where
$\hL^{(i)}=\hV_i^\dag[\id_\chi\otimes\hh\big]\hV_i$,
and we have defined the site-transition map (quantum channel)
\begin{equation}
	\M_t(\hR):=\sum_{s=1}^d \hV^s_t\hR \hV^{s\dag}_t.
\end{equation}
According to Eq.~\eqref{eq:gradRiemann}, the contribution $\Avg\Tr\big(\hg_\tau^{(i)\dag} \hg_\tau^{(i)}\big)$ to the gradient variance \eqref{eq:var0} is quadratic in both $\hX_\tau^{(i)}$ and $\hY_\tau^{(i)}$. The essential step is hence to evaluate the Haar averages $\Avg \hX_\tau^{(i)}\otimes \hX_\tau^{(i)}$ and $\Avg \hY_\tau^{(i)}\otimes\hY_\tau^{(i)}$, i.e.,
\begin{align}
	\label{eq:mps-XXavg}
	&\Avg\M^{\otimes 2}_{\tau+1}\circ\dotsb\circ\M^{\otimes 2}_{L}(|0,0\ket\bra 0,0|)\quad\text{and}\\
	\label{eq:mps-YYavg}
	&\Avg\M^{\dag\otimes 2}_{\tau-1}\circ\dotsb\circ\M^{\dag\otimes 2}_{i+1}(\hL^{(i)}\otimes\hL^{(i)}).
\end{align}
Taking the Haar average of $\M^{\otimes 2}_t$ over the corresponding unitary $\hU_t$ with $t\in\{\tau+1,\dotsc,L\}$ and $t\in\{i+1,\dotsc,\tau-1\}$, respectively, yields the doubled site-transition channel
\begin{equation}\label{eq:mps-E2}
	 \E_\mps^\doub :=\Avg_{\hU_t}\M^{\otimes 2}_t= |\hr_1\kket\bbra \id_{\chi^2}| + \eta_\mps\, |\hr_2\kket\bbra\hl_2|.
\end{equation}
Here, we have already written its diagonalized form, using a super-bra-ket notation for operators based on the Hilbert-Schmidt inner product $\bbra\hA|\hB\kket:=\Tr(\hA^\dag\hB)$. The left and right eigenvectors are biorthogonal, $\bbra\hl_i|\hr_j\kket= \delta_{i,j}$. The diagonalization shows that $\E_\mps^\doub$ only has the two nonzero eigenvalues 1 and $\eta_\mps=\frac{1-1/\chi^2}{d-1/(\chi^2 d)}$ with the corresponding unique steady state $\hr_1$ and the first excitation $\hr_2$. The repeated application of $\E_\mps^\doub$ in Eq.~\eqref{eq:mps-XXavg} quickly converges to $\hr_1$. Similarly, its application in Eq.~\eqref{eq:mps-YYavg} would converge to the corresponding left eigenoperator $\hl_1=\id_{\chi^2}$, but one finds that this does not contribute to the gradient variance. It is the subleading term $\propto \eta_\mps^{\tau-i}\hl_2$ that ultimately yields
\begin{equation}\label{eq:mps-var-g}
	 \frac{\Avg\Tr\big(\hg_\tau^{(i)\dag}\hg_\tau^{(i)}\big)}{\chi\, d}
	 =\frac{2\Tr(\hh^2)}{d(\chi^2d+1)}\, \eta_\mps^{\tau-i} + \O(\eta_\mps^{L-i}).
\end{equation}

Finally, the gradient variance \eqref{eq:var0} for the extensive Hamiltonian is obtained by summing the contributions \eqref{eq:mps-var-g} for all $i\leq \tau$, resulting in the system-size independent value
\begin{multline}\label{eq:mps-var-gTot1}
	\Var \Big(\partial_{\hU_\tau} \sum_i \bra\Psi|\hh_i|\Psi\ket\Big)
	=2\Tr(\hh^2)\frac{\chi^2d^2-1}{d(d-1)(\chi^2d+1)^2}\\  + \O(\eta_\mps^\tau) + \O(\eta_\mps^{L-\tau}), 
\end{multline}
where the sub-leading terms are due to boundary effects.
\begin{figure}[t]
	\label{fig:mps-var}
	\includegraphics[width=\columnwidth]{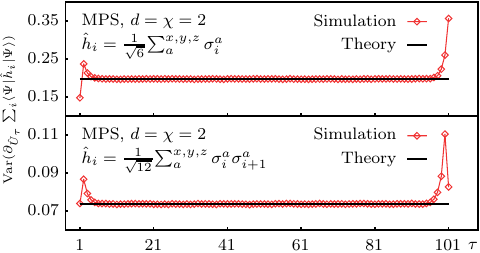}
	\caption{\textbf{MPS gradient variance.} For spin-1/2 chains \eqref{eq:H} of length $L=101$ with single-site terms $\hh_i=\sum_a\hsigma^a_i/\sqrt{6}$ (upper panel) and nearest-neighbor interactions $\hh_i=\sum_a\hsigma^a_i\hsigma^a_{i+1}/\sqrt{12}$ (lower panel), we plot the gradient variance \eqref{eq:var0} for the tensor at site $\tau$. Numerical averages over \numprint{64000} MPS with tensors sampled according to the Haar measure agree with the analytical results \eqref{eq:mps-var-gTot1} and \eqref{eq:mps-var-gTot2}.}
\end{figure}

The optimization problem with single-site terms $\hh_i$ is trivially solved by product states \cite{Note1}. In contrast, groundstate problems with finite-range interactions $\hh_i$ are Quantum-Merlin-Arthur complete \cite{Kempe2006-35,Oliveira2008-8,Aharonov2009-287,Gottesman2009_05,Bausch2017-18}. Fortunately, the analysis does not change qualitatively. The second largest eigenvalue $\eta_\mps$ of the doubled transition channel \eqref{eq:mps-E2} remains the most important quantity. Specifically, for nearest-neighbor interactions with $\hh_i=\id_d^{\otimes(i-1)}\otimes\hh\otimes\id_d^{\otimes(L-i-1)}$ acting non-trivially on sites $i$ and $i+1$, and assuming large $\chi$, we find
\begin{multline}\label{eq:mps-var-gTot2}
	\Var \Big(\partial_{\hU_\tau} \sum_i \bra\Psi|\hh_i|\Psi\ket\Big)
	\sim \frac{4}{\chi^2 d^4}\left[\Tr(\hh^2)+2\Tr(\Tr^2_1\hh)\right]\\ + \mc{O}(\eta_\mps^j) + \mc{O}(\eta_\mps^{L-j}), 
\end{multline}
A detailed derivation is given in the companion paper \cite{Barthel2023_03}. Figure~\ref{fig:mps-var} confirms the analytical prediction in numerical tests for spin-1/2 chains. Thus, MPS optimizations have \emph{no} barren plateaus for extensive Hamiltonians.

\begin{figure}[t]
	\label{fig:mps-var-scaleL}
	\includegraphics[width=\columnwidth]{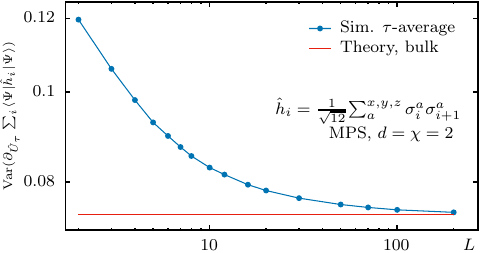}
	\caption{\textbf{Finite-size effects for MPS gradient variance.} The MPS energy-gradient variance
	for spin-1/2 Heisenberg antiferromagnets, averaged over all sites $\tau=1,\dotsc,L$, decays as a function of the chain length $L$ until converging to the predicted bulk value \eqref{eq:mps-var-gTot2}.}
\end{figure}

For open spin-1/2 Heisenberg chains and a subclass of MPS generated from single-qubit and CNOT gates, data in Ref.~\cite{Liu2019-1} suggests a power-law decay of the $\tau$-averaged energy-gradient variance when increasing the system size $L$. This decay turns out to be a finite-size effect: While the gradient variance for an MPS tensor $\hU_\tau$ with site $\tau$ in the bulk of the system has the (approximately) $\tau$-independent value \eqref{eq:mps-var-gTot2}, there are corrections for $\tau$ close to the system boundaries. The corrections decay exponentially in the distance from the system boundaries. For the small $L\leq 20$ analyzed in Ref.~\cite{Liu2019-1}, the boundary effects are still dominant and lead to the observed polynomial decay. Figure~\ref{fig:mps-var-scaleL} with $L=2,\dotsc,200$ shows that the initial decay of the $\tau$-averaged variance levels off at the predicted bulk value \eqref{eq:mps-var-gTot2} for larger $L$.

\section{Hierarchical TNS}
MERA \cite{Vidal-2005-12,Vidal2006,Barthel2010-105} are hierarchical TNS. Starting on $\N$ sites, in each renormalization step $\tau\to\tau+1$, we apply local unitary disentanglers $\hU_{\tau,k}^\dag$ of layer $\tau$ before the number of degrees of freedom is reduced by applying projections $\hV_{\tau,k}^\dag$, each mapping a group of $b$ sites into one renormalized site. The dimension $\chi$ for the Hilbert space of each renormalized site is the bond dimension, and $b$ is the so-called branching ratio. The process stops at the top layer $\tau=T$ by projecting each of the remaining $\N/b^{T}$ sites onto a reference state $|0_\chi\ket$.
The renormalization procedure, seen in reverse, prepares the MERA $|\Psi\ket$ starting in layer $\tau=T$ with the reference states $|0_\chi\ket$ and then proceeding down until reaching the physical layer $\tau=0$. TTNS \cite{Fannes1992-66,Otsuka1996-53,Shi2006-74,Murg2010-82,Tagliacozzo2009-80} are a subclass of MERA without disentanglers ($\hU_{\tau,k}=\id$).

As all tensors are isometric, the evaluation of local expectation values can be drastically simplified, because $\bra\Psi|\hh_i|\Psi\ket$ only depends on the tensors in the causal cone of $\hh_i$. See Fig.~\ref{fig:isoTNS}. In fact, the expectation value can again be written in a form very similar to Eq.~\eqref{eq:mps-hi-expect} but, now, we have transition maps $\M_{\tau,i}$ that map the causal-cone density operator $\dm_{\tau,i}$, representing the state on the $n_c$ renormalized sites in the causal cone after preparation steps $T\to T-1\to\dots\to\tau$,
into 
\begin{equation*}
	\dm_{\tau-1,i}=\M_{\tau,i}(\dm_{\tau,i})=\M_{\tau,i}\circ\dotsb\circ\M_{T,i}\left((|0_\chi\ket\bra 0_\chi|)^{\otimes n_c}\right).
\end{equation*}
Specifically, for the binary one-dimensional (1D) MERA in Fig.~\ref{fig:isoTNS}d and a three-site interaction term $\hh_i$, we start at the top layer with the ($n_c=3$)-site reference state $\dm_{T,i}=(|0_\chi\ket\bra 0_\chi|)^{\otimes 3}$ and then progress down layer by layer, applying either a left-moving or a right-moving transition map $\M_{t,i}$. These consist in applying three isometries $\hV_{t,k}\otimes\hV_{t,k+1}\otimes\hV_{t,k+2}$ that double the number of (renormalized) sites to six, then applying two disentanglers $\id_\chi\otimes\hU_{t,k}\otimes\hU_{t,k+1}\otimes\id_\chi$ and, finally, tracing out one site on the left and two on the right (left-moving) or vice versa (right-moving).

\begin{figure*}[p]
	\label{fig:mera-var}
	\centering
	\includegraphics[width=0.47\textwidth]{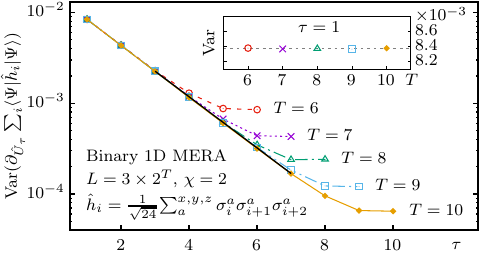}\hspace{0.03\textwidth}
	\includegraphics[width=0.47\textwidth]{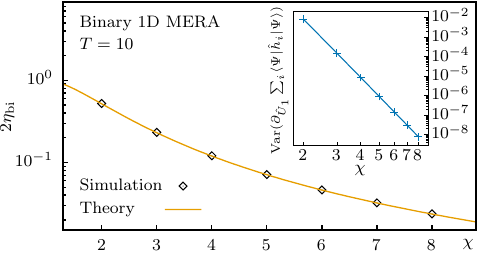}\\
	\vspace{1em}
	\includegraphics[width=0.47\textwidth]{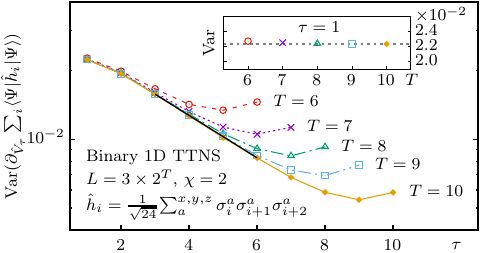}\hspace{0.03\textwidth}
	\includegraphics[width=0.47\textwidth]{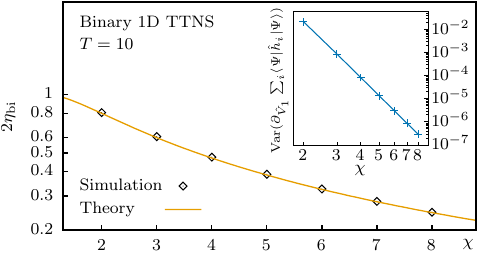}\\
	\vspace{1em}
	\includegraphics[width=0.47\textwidth]{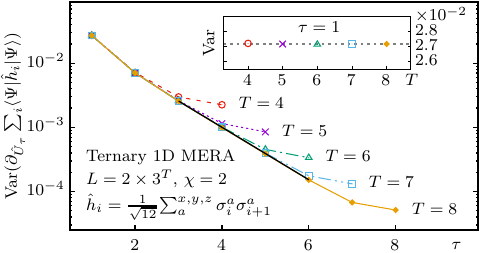}\hspace{0.03\textwidth}
	\includegraphics[width=0.47\textwidth]{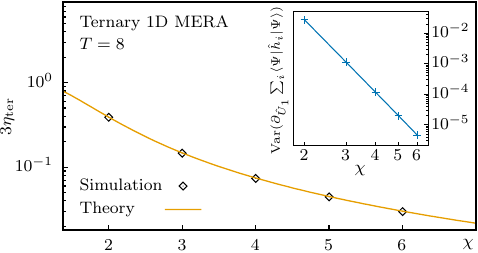}\\
	\vspace{1em}
	\includegraphics[width=0.47\textwidth]{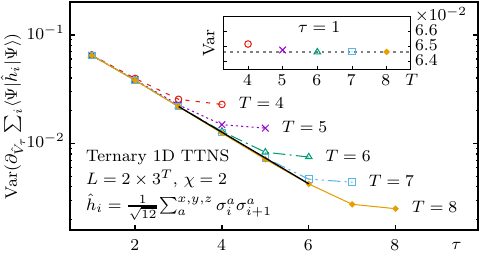}\hspace{0.03\textwidth}
	\includegraphics[width=0.47\textwidth]{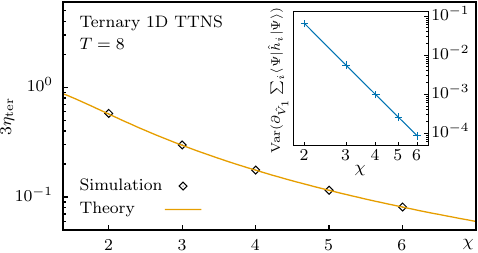}
	\caption{\textbf{Gradient variance in heterogeneous 1D MERA and TTNS.} In the hierarchical isometric TNS, the leading term in the Haar-variance of the energy gradient is system-size independent and decreases exponentially in the layer index $\tau$. The graphs on left show this exponential decay observed in numerical tests for bond dimension $\chi=2$. Black lines indicate the fitting range for the decay factors. The insets show the gradient variance at $\tau = 1$ as a function of the number of layers $T$, corresponding to varying system size $L\gtrsim b^T$. The graphs on the right show a perfect agreement of the decay factors, extracted from simulations for various bond dimensions $\chi$, and the theoretical prediction $b\eta$ according to Eqs.~\eqref{eq:MERAbin-E2-diag} and \eqref{eq:eta}. The insets assert the power-law decay of gradient variances in layer $\tau=1$ with respect to $\chi$. For simplicity, we choose the physical single-site dimensions as $d=\chi$ and Hamiltonians $\hH=\sum_{i=1}^L\hh_i$ with normalized two or three-site interactions $\hh_i$ constructed from generalized Gell-Mann matrices according to Eq.~\eqref{eq:GellMann-h}.}
\end{figure*}

The diagonal contributions to the gradient variance for $\hU_{\tau,k}$ with $i_1=i_2=i$ in Eq.~\eqref{eq:var0}, are linear functions of $\Avg \hX_{\tau,k}^{(i)}\otimes \hX_{\tau,k}^{(i)}$ and $\Avg \hY_{\tau,k}^{(i)}\otimes\hY_{\tau,k}^{(i)}$, where the additional label $k$ identifies the specific tensor in layer $\tau$. Taking the Haar average of $\M_{t,i}^{\otimes 2}$, we obtain either a left-moving or a right-moving doubled layer-transition channel $\E^\doub_{\bin,\tL}$ and $\E^\doub_{\bin,\tR}$. Summing over all sites $i\in\S_{\tau,k}$ that have $\hU_{\tau,k}$ in their causal cone, corresponds to summing over all possible sequences of the two channels. This is equivalent to applying the map $2\E^\doub_{\bin}$ for layers $t=1,\dotsc,\tau-1$, where
\begin{equation}\label{eq:MERAbin-E2}
	\E^\doub_\bin:= \frac{1}{2}\left(\E^\doub_{\bin,\tL}+\E^\doub_{\bin,\tR}\right)
\end{equation}
is the average transition channel. Finally, averaging the gradient variance $\Var\hg(\hU_{\tau,k})$ with respect to $k$ in layer $\tau$ corresponds to applying $\E^\doub_\bin$ for all layers $t=\tau+1,\dotsc,T$.

The channel $\E^\doub_\bin$ is diagonalizable and gapped,
\begin{align}\label{eq:MERAbin-E2-diag}
	&\E^\doub_\bin = |\hr_1\kket \bbra\hl_1| + \sum_{n=2}^4\lambda_n |\hr_n\kket \bbra\hl_n|\ \ \text{with}\\ \nonumber
	&
	\hl_1=\id_{\chi^6},\ \ 
	\eta_\bin := \lambda_2 = \frac{\chi^2(1+\chi)^4}{2(1+\chi^2)^4},\ \ 
	 \lambda_3 = \frac{\chi^2(1+\chi)^2}{2(1+\chi^2)^3},
\end{align}
biorthogonal left and right eigenvectors $\bbra\hl_n|\hr_{n'}\kket=\delta_{n,n'}$, and $\frac{1}{2}>\lambda_2>\lambda_3>\lambda_4$. Similar to the analysis for MPS, the leading term in $\Avg \hX_{\tau,k}^{(i)}\otimes \hX_{\tau,k}^{(i)}$ stems from the $\E^\doub_\bin$ steady state $\hr_1$, while the leading contributing term in $\Avg \hY_{\tau,k}^{(i)}\otimes \hX_{\tau,k}^{(i)}$ stems from the first excitation $\hl_2$. In this way, one finds that the diagonal contributions to the gradient variance $\Var\partial_{\hU_{\tau,k}}\bra\Psi|\hH|\Psi\ket$  [Eq.~\eqref{eq:var0}], averaged for all $k$ in layer $\tau$, scale as
\begin{equation}\label{eq:var-mera}
	\Var\hg(\hU_{\tau.k}) = \Theta((2\eta_\bin)^\tau) + \O((2\lambda_3)^\tau) + \O(2^\tau\eta_\bin^T),
\end{equation}
where the Landau symbol $\Theta(f)$ indicates that there exist upper and lower bounds scaling like $f$. The off-diagonal terms with $|i_1-i_2|>3$ vanish due to $\Tr\hh_i = 0$ and the remaining off-diagonal terms have the same scaling as the diagonal terms. See Ref.~\cite{Barthel2023_03} for details. 

The analysis for the binary 1D MERA can be extended to all MERA and TTNS. The central object in the evaluation of their Haar-averaged gradient variances are doubled layer-transition channels $\E^\doub$. The gradient variance for tensors in layer $\tau$ will then scale as $(b\eta)^\tau$, where $\eta$ is the second largest $\E^\doub$ eigenvalue. This eigenvalue decreases algebraically with increasing bond dimension $\chi$.
Specifically, we find
\begin{subequations}\label{eq:eta}
\begin{alignat}{3}
	&\eta&&=\frac{\chi}{1+\chi^2}\ \ &&\text{for binary 1D TTNS},\\
	&\eta&&=\frac{1}{3\chi^2}+\O(\chi^{-4})\ \ &&\text{for ternary 1D MERA},\\
	&\eta&&=\frac{\chi^2}{1+\chi^2+\chi^4}\ \ &&\text{for ternary 1D TTNS, and}\\
	&\eta&&=\frac{1}{9\chi^8}+\O(\chi^{-10})\ \ &&\text{for nonary 2D MERA \cite{Evenbly2009-79}}.
\end{alignat}
\end{subequations}
So, for each layer $\tau$, the gradient variance is an algebraic function of the bond dimension $\chi$ and, up to subleading corrections, independent of the total system size. Therefore, the optimization of hierarchical TNS is \emph{not} hampered by barren plateaus.

For the 1D hierarchical TNS and Hamiltonians with two-site and three-site interactions $\hh_i$, these analytical results are tested and confirmed numerically as shown in Fig.~\ref{fig:mera-var}. For the numerical tests, we choose the physical single-site dimension $d$ equal to the bond dimension $\chi$. Otherwise, one would use the lowest MERA layers to increase bond dimensions gradually from $d$ to the desired $\chi$. The isotropic interaction terms $\hh_i$ are constructed using generalized $\chi\times\chi$ Gell-Mann matrices $\{\hLambda^{1},\hLambda^{2},\cdots,\hLambda^{\chi^2-1}\}$ \cite{Kimura2003-314,Bertlmann2008-41}. They are traceless and Hermitian generalizations of the Pauli matrices $\hsigma^a$ for $\chi=2$ and the Gell-Mann matrices for $\chi=3$. As generators of the special unitary group SU$(\chi)$, they satisfy the orthonormality condition $\bbra\hLambda^a |\hLambda^b\kket = 2\delta_{a,b}$. We define $m$-site interactions as
\begin{equation}\label{eq:GellMann-h}
	\hh_i = \frac{1}{\sqrt{2^m(\chi^2-1)}}\sum_{a=1}^{\chi^2-1} \hLambda^a_{i}\otimes\dotsb\otimes\hLambda^a_{i+m-1},
\end{equation}
which are traceless, have vanishing partial traces, and are normalized according to $\Tr(\hh_i^2)=1$. Each data point in Fig.~\ref{fig:mera-var} corresponds to 1000 TNS with all tensors sampled according the Haar measure. The numerical results confirm the scaling $(b\eta)^\tau$ with corrections at small $\tau$ and small $T-\tau$. The extracted decay factors $b\eta$ display the expected $\chi$ dependence in accordance with Eqs.~\eqref{eq:MERAbin-E2-diag} and \eqref{eq:eta}.
As a numerically sampling of 2D MERA and TTNS is computationally expensive, a direct extraction of decay factors from sampled gradient variances is currently not possible for 2D.

\section{Homogeneous and Trotterized MERA and TTNS}
\begin{figure}[t]
	\label{fig:tmera-var}
	\includegraphics[width=\columnwidth]{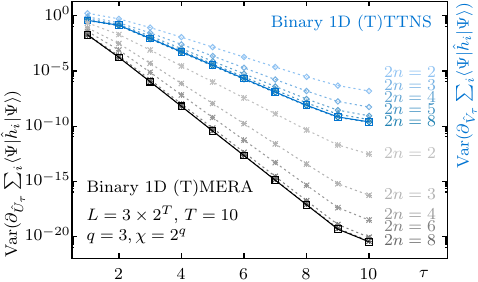}
	\caption{\textbf{Gradient variances for homogeneous Trotterized TNS.} For the model \eqref{eq:GellMann-h}, the plot shows the gradient variances at layer $\tau$ for homogeneous binary 1D Trotterized TTNS (blue dashed lines), full TTNS (blue full line), Trotterized MERA (gray dashed lines) and full MERA (black full line), where $n$ denotes the number of Trotter steps per tensor \cite{Miao2021_08,Miao2023_03}.}
\end{figure}
So far, we have exclusively considered heterogeneous TNS, where all tensors can vary freely. To save computational resources, one can also work with homogeneous TNS where, in the case of MERA and TTNS, all equivalent disentanglers and isometries of a given layer $\tau$ are chosen to be identical. In the case of the binary MERA, we can for example set $\hU_{\tau,k}\equiv\hU_\tau$ and $\hV_{\tau,k}\equiv\hV_\tau$ for all disentanglers and isometries in layer $\tau$. Such homogeneous TNS can also be used to initialized the optimization of heterogeneous TNS. The theoretical analysis of gradient variances for homogeneous TNS is more involved. While the analysis for heterogeneous TNS only requires first and second-moment Haar-measure integrals, more complicated higher-order integrals are needed for the homogeneous states. Hence, we determine the gradient variance for homogeneous TTNS and MERA numerically as shown in Fig.~\ref{fig:tmera-var}, finding that homogeneous TNS have considerably larger gradient variances than the corresponding heterogeneous states. This is consistent with findings in Refs.~\cite{Volkoff2021-6,Pesah2021-11} for other classes of states.

Isometric TNS can be implemented on quantum computers, but it is advisable to impose a substructure for the TNS tensors to reduce costs and achieve a quantum advantage. Specifically, in Trotterized MERA \cite{Miao2021_08,Miao2023_03,Kim2017_11,Haghshenas2022-12,Haghshenas2023_05}, each tensor is constructed as brickwall circuit with $n$ (Trotter) steps. A generic full MERA can be recovered by increasing $n$. Figure~\ref{fig:tmera-var} compares gradient variances for homogeneous Trotterized MERA and full MERA as well as those for Trotterized and full TTNS. The data for three-site interactions \eqref{eq:GellMann-h} shows that Trotterized TNS feature larger gradient variances than full TNS, and the former converge to the latter as the number $n$ of Trotter steps increases.

\section{Discussion}
The presented results substantiate that the considered isometric TNS generally feature \emph{no} barren plateaus in the energy optimization for extensive models with finite-range interactions.
This opens a route to efficiently solve groundstate problems for strongly-correlated condensed matter systems with VQA on small quantum computers, e.g., using Trotterized MERA \cite{Miao2021_08,Miao2023_03,Haghshenas2022-12}. For this approach, the benchmark data in Refs.~\cite{Miao2021_08,Miao2023_03} implies polynomial quantum advantages which will strongly increase with increasing number of spatial dimensions. In this way, limitations due to high classical tensor contractions costs could be overcome.
First experiments demonstrated critical correlations in preoptimized Trotterized 1D MERA \cite{Haghshenas2023_05}.

The observed scaling of the gradient variance has implications for efficient initialization schemes in quantum and classical algorithms: For MPS, the power-law decay in the bond dimension $\chi$ suggests to start with an optimization at small $\chi$ and to then gradually increase it. For TTNS and MERA, the exponential decay in the layer index $\tau$, suggests that iteratively increasing the number of layers during optimization can substantially improve the performance. TTNS have considerably larger gradient variances than MERA. For MERA optimizations, it can hence be beneficial to initially choose all disentanglers as identities and only start their optimization after the corresponding TTNS has converged.

For VQA with broader classes of quantum circuits, our results suggests that isometric TNS might be a good starting point. One can start with a well-optimized TNS and, subsequently, gradually introduce further quantum gates, which can extend the expressiveness beyond what is achievable in classical simulations without compromising trainability \cite{Dborin2022-7,Rudolph2023-14}.

The analysis in Ref.~\cite{Kim2017_11} suggests that VQA based on hierarchical TNS is robust with respect to noise in the quantum devices. A natural extension of our analysis would be to study the doubled transition channels with environment fluctuations. We expect that no noise-induced barren plateaus \cite{Wang2021-12} occur for TNS.

Our results can be generalized to systems with arbitrary finite-range and, more generally, $k$-local interactions.
The proof technique employed in this work, which is centered around the analysis of the doubled (layer or site) transition quantum channels, could also be applied to study 
statistical properties and typicality for random TNS \cite{Garnerone2010-81,Garnerone2010-82,Haferkamp2021-2}, as well as the dynamics of quantum information and entanglement in structured random quantum circuits~\cite{Nahum2017-7,Nahum2018-8,Zhou2019-99,Potter2022-211,Fisher2023-14}.

The absence of barren plateaus in the discussed isometric TNS does not depend on whether one employs Riemannian gradients or Euclidean gradients. However, Riemannian gradients simplify the analytical analysis and, in our practical experience, the corresponding parametrization-free Riemannian optimization as described in Ref.~\cite{Miao2021_08} has better convergence properties, also mitigating some effects of spurious local stationary points observed for parametrized quantum circuits~\cite{You2021-139,Anschuetz2022-13,Liu2022_06}. Further research on this issues is needed.\vspace{1em}

\begin{acknowledgments}
We gratefully acknowledge discussions with Kenneth R.\ Brown, Daniel Stilck Fran\c{c}a, Jin-Guo Liu, and Iman Marvian as well as support through US Department of Energy grant DE-SC0019449 and the US National Science Foundation (NSF) Quantum Leap Challenge Institute for Robust Quantum Simulation (Award No.\ OMA-2120757).
\end{acknowledgments}

\appendix

\bibliographystyle{prsty.tb.title}

\begin{thebibliography}{10}

\bibitem{Feynman1982-21}
R.~P. Feynman, {\em Simulating physics with computers},
  \href{https://doi.org/10.1007/BF02650179} {Int. J. Theor. Phys. {\bf 21},
  467  (1982)}.

\bibitem{Cerezo2021-3}
M. Cerezo, A. Arrasmith, R. Babbush, S.~C. Benjamin, S. Endo, K. Fujii, J.~R.
  McClean, K. Mitarai, X. Yuan, L. Cincio, and P.~J. Coles, {\em Variational
  quantum algorithms}, \href{https://doi.org/10.1038/s42254-021-00348-9} {Nat.
  Rev. Phys. {\bf 3},  625  (2021)}.

\bibitem{Kiani2020_01}
B.~T. Kiani, S. Lloyd, and R. Maity, {\em Learning unitaries by gradient
  descent}, \href{http://arxiv.org/abs/2001.11897} {arXiv:2001.11897  (2020)}.

\bibitem{Bittel2021-127}
L. Bittel and M. Kliesch, {\em Training variational quantum algorithms is
  NP-hard}, \href{https://doi.org/10.1103/PhysRevLett.127.120502} {Phys. Rev.
  Lett. {\bf 127},  120502  (2021)}.

\bibitem{Anschuetz2021_09}
E.~R. Anschuetz, {\em Critical points in quantum generative models},
  \href{http://arxiv.org/abs/2109.06957} {arXiv:2109.06957  (2021)}.

\bibitem{Anschuetz2022-13}
E.~R. Anschuetz and B.~T. Kiani, {\em Quantum variational algorithms are
  swamped with traps}, \href{https://doi.org/10.1038/s41467-022-35364-5} {Nat.
  Commun. {\bf 13},  7760  (2022)}.

\bibitem{McClean2018-9}
J.~R. McClean, S. Boixo, V.~N. Smelyanskiy, R. Babbush, and H. Neven, {\em
  Barren plateaus in quantum neural network training landscapes},
  \href{https://doi.org/10.1038/s41467-018-07090-4} {Nat. Commun. {\bf 9},
  4812  (2018)}.

\bibitem{Cerezo2021-12}
M. Cerezo, A. Sone, T. Volkoff, L. Cincio, and P.~J. Coles, {\em Cost function
  dependent barren plateaus in shallow parametrized quantum circuits},
  \href{https://doi.org/10.1038/s41467-021-21728-w} {Nat. Commun. {\bf 12},
  1791  (2021)}.

\bibitem{Arrasmith2022-7}
A. Arrasmith, Z. Holmes, M. Cerezo, and P.~J. Coles, {\em Equivalence of
  quantum barren plateaus to cost concentration and narrow gorges},
  \href{https://doi.org/10.1088/2058-9565/ac7d06} {Quantum Sci. Technol. {\bf
  7},  045015  (2022)}.

\bibitem{Miao2024_02}
Q. Miao and T. Barthel, {\em Equivalence of cost concentration and gradient
  vanishing for quantum circuits: An elementary proof in the Riemannian
  formulation}, \href{http://arxiv.org/abs/2402.07883} {arXiv:2402.07883
  (2024)}.

\bibitem{Grant2019-3}
E. Grant, L. Wossnig, M. Ostaszewski, and M. Benedetti, {\em An initialization
  strategy for addressing barren plateaus in parametrized quantum circuits},
  \href{https://doi.org/10.22331/q-2019-12-09-214} {Quantum {\bf 3},  214
  (2019)}.

\bibitem{Zhang2022_03}
K. Zhang, M.-H. Hsieh, L. Liu, and D. Tao, {\em Gaussian initializations help
  deep variational quantum circuits escape from the barren plateau},
  \href{http://arxiv.org/abs/2203.09376} {arXiv:2203.09376  (2022)}.

\bibitem{Mele2022-106}
A.~A. Mele, G.~B. Mbeng, G.~E. Santoro, M. Collura, and P. Torta, {\em Avoiding
  barren plateaus via transferability of smooth solutions in a Hamiltonian
  variational ansatz}, \href{https://doi.org/10.1103/PhysRevA.106.L060401}
  {Phys. Rev. A {\bf 106},  L060401  (2022)}.

\bibitem{Kulshrestha2022_04}
A. Kulshrestha and I. Safro, {\em BEINIT: Avoiding barren plateaus in
  variational quantum algorithms}, \href{http://arxiv.org/abs/2204.13751}
  {arXiv:2204.13751  (2022)}.

\bibitem{Dborin2022-7}
J. Dborin, F. Barratt, V. Wimalaweera, L. Wright, and A.~G. Green, {\em Matrix
  product state pre-training for quantum machine learning},
  \href{https://doi.org/10.1088/2058-9565/ac7073} {Quant. Sci. Tech. {\bf 7},
  035014  (2022)}.

\bibitem{Skolik2021-3}
A. Skolik, J.~R. McClean, M. Mohseni, P. van~der Smagt, and M. Leib, {\em
  Layerwise learning for quantum neural networks},
  \href{https://doi.org/10.1007/s42484-020-00036-4} {Quantum Mach. Intell. {\bf
  3},  5  (2021)}.

\bibitem{Slattery2022-4}
L. Slattery, B. Villalonga, and B.~K. Clark, {\em Unitary block optimization
  for variational quantum algorithms},
  \href{https://doi.org/10.1103/PhysRevResearch.4.023072} {Phys. Rev. Research
  {\bf 4},  023072  (2022)}.

\bibitem{Haug2021_04}
T. Haug and M. Kim, {\em Optimal training of variational quantum algorithms
  without barren plateaus}, \href{http://arxiv.org/abs/2104.14543}
  {arXiv:2104.14543  (2021)}.

\bibitem{Sack2022-3}
S.~H. Sack, R.~A. Medina, A.~A. Michailidis, R. Kueng, and M. Serbyn, {\em
  Avoiding barren plateaus using classical shadows},
  \href{https://doi.org/10.1103/PRXQuantum.3.020365} {PRX Quantum {\bf 3},
  020365  (2022)}.

\bibitem{Rad2022_03}
A. Rad, A. Seif, and N.~M. Linke, {\em Surviving the barren plateau in
  variational quantum circuits with Bayesian learning initialization},
  \href{http://arxiv.org/abs/2203.02464} {arXiv:2203.02464  (2022)}.

\bibitem{Tao2022_05}
Z. Tao, J. Wu, Q. Xia, and Q. Li, {\em LAWS: Look around and warm-start natural
  gradient descent for quantum neural networks},
  \href{http://arxiv.org/abs/2205.02666} {arXiv:2205.02666  (2022)}.

\bibitem{Campos2021-103}
E. Campos, A. Nasrallah, and J. Biamonte, {\em Abrupt transitions in
  variational quantum circuit training},
  \href{https://doi.org/10.1103/PhysRevA.103.032607} {Phys. Rev. A {\bf 103},
  032607  (2021)}.

\bibitem{Dankert2009-80}
C. Dankert, R. Cleve, J. Emerson, and E. Livine, {\em Exact and approximate
  unitary 2-designs and their application to fidelity estimation},
  \href{https://doi.org/10.1103/PhysRevA.80.012304} {Phys. Rev. A {\bf 80},
  012304  (2009)}.

\bibitem{Brandao2016-346}
F.~G. S.~L. Brand{\~a}o, A.~W. Harrow, and M. Horodecki, {\em Local random
  quantum circuits are approximate polynomial-designs},
  \href{https://doi.org/10.1007/s00220-016-2706-8} {Commun. Math. Phys. {\bf
  346},  397  (2016)}.

\bibitem{Harrow2023-05}
A.~W. Harrow and S. Mehraban, {\em Approximate unitary $t$-designs by short
  random quantum circuits using nearest-neighbor and long-range gates},
  \href{https://doi.org/10.1007/s00220-023-04675-z} {Commun. Math. Phys.
  (2023)}.

\bibitem{Sim2019-2}
S. Sim, P.~D. Johnson, and A. Aspuru-Guzik, {\em Expressibility and entangling
  capability of parameterized quantum circuits for hybrid quantum-classical
  algorithms}, \href{https://doi.org/10.1002/qute.201900070} {Adv. Quantum
  Technol. {\bf 2},  1900070  (2019)}.

\bibitem{Nakaji2021-5}
K. Nakaji and N. Yamamoto, {\em Expressibility of the alternating layered
  ansatz for quantum computation},
  \href{https://doi.org/10.22331/q-2021-04-19-434} {Quantum {\bf 5},  434
  (2021)}.

\bibitem{Du2022-128}
Y. Du, Z. Tu, X. Yuan, and D. Tao, {\em Efficient measure for the expressivity
  of variational quantum algorithms},
  \href{https://doi.org/10.1103/PhysRevLett.128.080506} {Phys. Rev. Lett. {\bf
  128},  080506  (2022)}.

\bibitem{Holmes2022-3}
Z. Holmes, K. Sharma, M. Cerezo, and P.~J. Coles, {\em Connecting ansatz
  expressibility to gradient magnitudes and barren plateaus},
  \href{https://doi.org/10.1103/PRXQuantum.3.010313} {PRX Quantum {\bf 3},
  010313  (2022)}.

\bibitem{Goldstein2006-96}
S. Goldstein, J.~L. Lebowitz, R. Tumulka, and N. Zangh\`{\i}, {\em Canonical
  typicality}, \href{https://doi.org/10.1103/PhysRevLett.96.050403} {Phys. Rev.
  Lett. {\bf 96},  050403  (2006)}.

\bibitem{Popescu2006-2}
S. Popescu, A.~J. Short, and A. Winter, {\em Entanglement and the foundations
  of statistical mechanics}, \href{https://doi.org/doi:10.1038/nphys444} {Nat.
  Phys. {\bf 2},  754  (2006)}.

\bibitem{Ortiz2021-2}
C. Ortiz~Marrero, M. Kieferov\'a, and N. Wiebe, {\em Entanglement-induced
  barren plateaus}, \href{https://doi.org/10.1103/PRXQuantum.2.040316} {PRX
  Quantum {\bf 2},  040316  (2021)}.

\bibitem{Patti2021-3}
T.~L. Patti, K. Najafi, X. Gao, and S.~F. Yelin, {\em Entanglement devised
  barren plateau mitigation},
  \href{https://doi.org/10.1103/PhysRevResearch.3.033090} {Phys. Rev. Research
  {\bf 3},  033090  (2021)}.

\bibitem{Sharma2022-128}
K. Sharma, M. Cerezo, L. Cincio, and P.~J. Coles, {\em Trainability of
  dissipative perceptron-based quantum neural networks},
  \href{https://doi.org/10.1103/PhysRevLett.128.180505} {Phys. Rev. Lett. {\bf
  128},  180505  (2022)}.

\bibitem{Schollwoeck2011-326}
U. Schollw\"{o}ck, {\em The density-matrix renormalization group in the age of
  matrix product states}, \href{https://doi.org/10.1016/j.aop.2010.09.012}
  {Ann. Phys. {\bf 326},  96  (2011)}.

\bibitem{Barthel2022-112}
T. Barthel, J. Lu, and G. Friesecke, {\em On the closedness and geometry of
  tensor network state sets}, \href{https://doi.org/10.1007/s11005-022-01552-z}
  {Lett. Math. Phys. {\bf 112},  72  (2022)}.

\bibitem{Baxter1968-9}
R.~J. Baxter, {\em Dimers on a rectangular lattice},
  \href{https://doi.org/10.1063/1.1664623} {J. Math. Phys. {\bf 9},  650
  (1968)}.

\bibitem{Accardi1981}
L. Accardi, {\em Topics in quantum probability},
  \href{https://doi.org/10.1016/0370-1573(81)90070-3} {Phys. Rep. {\bf 77},
  169  (1981)}.

\bibitem{Fannes1992-144}
M. Fannes, B. Nachtergaele, and R.~F. Werner, {\em Finitely correlated states
  on quantum spin chains}, \href{https://doi.org/10.1007/BF02099178} {Commun.
  Math. Phys. {\bf 144},  443  (1992)}.

\bibitem{White1992-11}
S.~R. White, {\em Density matrix formulation for quantum renormalization
  groups}, \href{https://doi.org/10.1103/PhysRevLett.69.2863} {Phys. Rev. Lett.
  {\bf 69},  2863  (1992)}.

\bibitem{Rommer1997}
S. Rommer and S. \"Ostlund, {\em A class of ansatz wave functions for 1D spin
  systems and their relation to DMRG},
  \href{https://doi.org/10.1103/PhysRevB.55.2164} {Phys. Rev. B {\bf 55},  2164
   (1997)}.

\bibitem{PerezGarcia2007-7}
D. Perez-Garcia, F. Verstraete, M.~M. Wolf, and J.~I. Cirac, {\em Matrix
  product state representations}, {Quantum Info. Comput. {\bf 7},  401
  (2007)}.

\bibitem{Fannes1992-66}
M. Fannes, B. Nachtergaele, and R.~F. Werner, {\em Ground states of {VBS}
  models on cayley trees}, \href{https://doi.org/10.1007/bf01055710} {J. Stat.
  Phys. {\bf 66},  939  (1992)}.

\bibitem{Otsuka1996-53}
H. Otsuka, {\em Density-matrix renormalization-group study of the spin-$1/2$
  $\mathrm{XXZ}$ antiferromagnet on the Bethe lattice},
  \href{https://doi.org/10.1103/PhysRevB.53.14004} {Phys. Rev. B {\bf 53},
  14004  (1996)}.

\bibitem{Shi2006-74}
Y.-Y. Shi, L.-M. Duan, and G. Vidal, {\em Classical simulation of quantum
  many-body systems with a tree tensor network},
  \href{https://doi.org/10.1103/PhysRevA.74.022320} {Phys. Rev. A {\bf 74},
  022320  (2006)}.

\bibitem{Murg2010-82}
V. Murg, F. Verstraete, O. Legeza, and R.~M. Noack, {\em Simulating strongly
  correlated quantum systems with tree tensor networks},
  \href{https://doi.org/10.1103/PhysRevB.82.205105} {Phys. Rev. B {\bf 82},
  205105  (2010)}.

\bibitem{Tagliacozzo2009-80}
L. Tagliacozzo, G. Evenbly, and G. Vidal, {\em Simulation of two-dimensional
  quantum systems using a tree tensor network that exploits the entropic area
  law}, \href{https://doi.org/10.1103/PhysRevB.80.235127} {Phys. Rev. B {\bf
  80},  235127  (2009)}.

\bibitem{Vidal-2005-12}
G. Vidal, {\em Entanglement renormalization},
  \href{https://doi.org/10.1103/PhysRevLett.99.220405} {Phys. Rev. Lett. {\bf
  99},  220405  (2007)}.

\bibitem{Vidal2006}
G. Vidal, {\em Class of quantum many-body states that can be efficiently
  simulated}, \href{https://doi.org/10.1103/PhysRevLett.101.110501} {Phys. Rev.
  Lett. {\bf 101},  110501  (2008)}.

\bibitem{Barthel2010-105}
T. Barthel, M. Kliesch, and J. Eisert, {\em Real-space renormalization yields
  finitely correlated states},
  \href{https://doi.org/10.1103/PhysRevLett.105.010502} {Phys. Rev. Lett. {\bf
  105},  010502  (2010)}.

\bibitem{Evenbly2022-8}
G. Evenbly, {\em A practical guide to the numerical implementation of tensor
  networks I: Contractions, decompositions, and gauge freedom},
  \href{https://doi.org/10.3389/fams.2022.806549} {Front. Appl. Math. Stat.
  {\bf 8},    (2022)}.

\bibitem{Miao2021_08}
Q. Miao and T. Barthel, {\em Quantum-classical eigensolver using multiscale
  entanglement renormalization},
  \href{https://doi.org/10.1103/PhysRevResearch.5.033141} {Phys. Rev. Research
  {\bf 5},  033141  (2023)}.

\bibitem{Liu2019-1}
J.-G. Liu, Y.-H. Zhang, Y. Wan, and L. Wang, {\em Variational quantum
  eigensolver with fewer qubits},
  \href{https://doi.org/10.1103/PhysRevResearch.1.023025} {Phys. Rev. Research
  {\bf 1},  023025  (2019)}.

\bibitem{FossFeig2021-3}
M. Foss-Feig, D. Hayes, J.~M. Dreiling, C. Figgatt, J.~P. Gaebler, S.~A. Moses,
  J.~M. Pino, and A.~C. Potter, {\em Holographic quantum algorithms for
  simulating correlated spin systems},
  \href{https://doi.org/10.1103/PhysRevResearch.3.033002} {Phys. Rev. Research
  {\bf 3},  033002  (2021)}.

\bibitem{Niu2022-3}
D. Niu, R. Haghshenas, Y. Zhang, M. Foss-Feig, G.~K.-L. Chan, and A.~C. Potter,
  {\em Holographic simulation of correlated electrons on a trapped-ion quantum
  processor}, \href{https://doi.org/10.1103/PRXQuantum.3.030317} {PRX Quantum
  {\bf 3},  030317  (2022)}.

\bibitem{Slattery2021_08}
L. Slattery and B.~K. Clark, {\em Quantum circuits for two-dimensional
  isometric tensor networks}, \href{http://arxiv.org/abs/2108.02792}
  {arXiv:2108.02792  (2021)}.

\bibitem{Miao2023_03}
Q. Miao and T. Barthel, {\em Convergence and quantum advantage of Trotterized
  MERA for strongly-correlated systems}, \href{http://arxiv.org/abs/2303.08910}
  {arXiv:2303.08910  (2023)}.

\bibitem{Liu2022-129}
Z. Liu, L.-W. Yu, L.-M. Duan, and D.-L. Deng, {\em Presence and absence of
  barren plateaus in tensor-network based machine learning},
  \href{https://doi.org/10.1103/PhysRevLett.129.270501} {Phys. Rev. Lett. {\bf
  129},  270501  (2022)}.

\bibitem{Garcia2023-2023}
R.~J. Garcia, C. Zhao, K. Bu, and A. Jaffe, {\em Barren plateaus from learning
  scramblers with local cost functions},
  \href{https://doi.org/10.1007/JHEP01(2023)090} {J. High Energ. Phys. {\bf
  2023},  90  (2023)}.

\bibitem{Zhao2021-5}
C. Zhao and X.-S. Gao, {\em Analyzing the barren plateau phenomenon in training
  quantum neural networks with the {ZX}-calculus},
  \href{https://doi.org/10.22331/q-2021-06-04-466} {Quantum {\bf 5},  466
  (2021)}.

\bibitem{Martin2023-7}
E. Cervero~Mart{\'{i}}n, K. Plekhanov, and M. Lubasch, {\em Barren plateaus in
  quantum tensor network optimization},
  \href{https://doi.org/10.22331/q-2023-04-13-974} {Quantum {\bf 7},  974
  (2023)}.

\bibitem{Note1}
For a system of $L$ sites, the expectation value $\bra \Psi |\hh _1\otimes \hh
  _2\otimes \dotsb \otimes \hh _L|\Psi \ket $ for normalized states $\Psi $ is
  simply minimized by the tensor product $|\Psi \ket =|\psi _1\ket \otimes
  \dotsb \otimes |\psi _L\ket $ of the lowest-eigenvalue eigenstates $\psi _i$
  of the single-site Hamiltonians $\hat {h}_i$. The same holds for sums $\bra
  \Psi |\sum _{i=1}^L\hh _i|\Psi \ket $ of single-site Hamiltonians.

\bibitem{Kempe2006-35}
J. Kempe, A. Kitaev, and O. Regev, {\em The complexity of the local Hamiltonian
  problem}, \href{https://doi.org/10.1137/S0097539704445226} {SIAM J. Comput.
  {\bf 35},  1070  (2006)}.

\bibitem{Oliveira2008-8}
R. Oliveira and B.~M. Terhal, {\em The complexity of quantum spin systems on a
  two-dimensional square lattice}, {Quantum Info. Comput. {\bf 8},  0900
  (2008)}.

\bibitem{Aharonov2009-287}
D. Aharonov, D. Gottesman, S. Irani, and J. Kempe, {\em The power of quantum
  systems on a line}, \href{https://doi.org/10.1007/s00220-008-0710-3} {Commun.
  Math. Phys. {\bf 287},  41  (2009)}.

\bibitem{Gottesman2009_05}
D. Gottesman and S. Irani, {\em The quantum and classical complexity of
  translationally invariant tiling and Hamiltonian problems},
  \href{https://doi.org/10.4086/toc.2013.v009a002} {Theory of Computing {\bf
  9},  31  (2013)}.

\bibitem{Bausch2017-18}
J. Bausch, T. Cubitt, and M. Ozols, {\em The complexity of translationally
  invariant spin chains with low local dimension},
  \href{https://doi.org/10.1007/s00023-017-0609-7} {Ann. Henri Poincar{\'e}
  {\bf 18},  3449  (2017)}.

\bibitem{Barthel2023_03}
T. Barthel and Q. Miao, {\em Absence of barren plateaus and scaling of
  gradients in the energy optimization of isometric tensor network states},
  \href{http://arxiv.org/abs/2304.00161} {arXiv:2304.00161  (2023)}.

\bibitem{Hauru2021-10}
M. Hauru, M. Van~Damme, and J. Haegeman, {\em Riemannian optimization of
  isometric tensor networks},
  \href{https://doi.org/10.21468/scipostphys.10.2.040} {SciPost Phys. {\bf 10},
     (2021)}.

\bibitem{Luchnikov2021-23}
I.~A. Luchnikov, M.~E. Krechetov, and S.~N. Filippov, {\em Riemannian geometry
  and automatic differentiation for optimization problems of quantum physics
  and quantum technologies}, \href{https://doi.org/10.1088/1367-2630/ac0b02}
  {New J. Phys. {\bf 23},  073006  (2021)}.

\bibitem{Wiersema2023-107}
R. Wiersema and N. Killoran, {\em Optimizing quantum circuits with Riemannian
  gradient flow}, \href{https://doi.org/10.1103/PhysRevA.107.062421} {Phys.
  Rev. A {\bf 107},  062421  (2023)}.

\bibitem{Evenbly2009-79}
G. Evenbly and G. Vidal, {\em Algorithms for entanglement renormalization},
  \href{https://doi.org/10.1103/PhysRevB.79.144108} {Phys. Rev. B {\bf 79},
  144108  (2009)}.

\bibitem{Kimura2003-314}
G. Kimura, {\em The Bloch vector for N-level systems},
  \href{https://doi.org/10.1016/s0375-9601(03)00941-1} {Phys. Lett. A {\bf
  314},  339  (2003)}.

\bibitem{Bertlmann2008-41}
R.~A. Bertlmann and P. Krammer, {\em Bloch vectors for qudits},
  \href{https://doi.org/10.1088/1751-8113/41/23/235303} {J. Phys. A: Math.
  Theor. {\bf 41},  235303  (2008)}.

\bibitem{Volkoff2021-6}
T. Volkoff and P.~J. Coles, {\em Large gradients via correlation in random
  parameterized quantum circuits},
  \href{https://doi.org/10.1088/2058-9565/abd891} {Quantum Sci. Technol. {\bf
  6},  025008  (2021)}.

\bibitem{Pesah2021-11}
A. Pesah, M. Cerezo, S. Wang, T. Volkoff, A.~T. Sornborger, and P.~J. Coles,
  {\em Absence of barren plateaus in quantum convolutional neural networks},
  \href{https://doi.org/10.1103/PhysRevX.11.041011} {Phys. Rev. X {\bf 11},
  041011  (2021)}.

\bibitem{Kim2017_11}
I.~H. Kim and B. Swingle, {\em Robust entanglement renormalization on a noisy
  quantum computer}, \href{http://arxiv.org/abs/1711.07500} {arXiv:1711.07500
  (2017)}.

\bibitem{Haghshenas2022-12}
R. Haghshenas, J. Gray, A.~C. Potter, and G.~K.-L. Chan, {\em Variational power
  of quantum circuit tensor networks},
  \href{https://doi.org/10.1103/PhysRevX.12.011047} {Phys. Rev. X {\bf 12},
  011047  (2022)}.

\bibitem{Haghshenas2023_05}
R. Haghshenas, E. Chertkov, M. DeCross, T.~M. Gatterman, J.~A. Gerber, K.
  Gilmore, D. Gresh, N. Hewitt, C.~V. Horst, M. Matheny, T. Mengle, B.
  Neyenhuis, D. Hayes, and M. Foss-Feig, {\em Probing critical states of matter
  on a digital quantum computer}, \href{http://arxiv.org/abs/2305.01650}
  {arXiv:2305.01650  (2023)}.

\bibitem{Rudolph2023-14}
M.~S. Rudolph, J. Miller, D. Motlagh, J. Chen, A. Acharya, and A.
  Perdomo-Ortiz, {\em Synergistic pretraining of parametrized quantum circuits
  via tensor networks}, \href{https://doi.org/10.1038/s41467-023-43908-6} {Nat.
  Commun. {\bf 14},    (2023)}.

\bibitem{Wang2021-12}
S. Wang, E. Fontana, M. Cerezo, K. Sharma, A. Sone, L. Cincio, and P.~J. Coles,
  {\em Noise-induced barren plateaus in variational quantum algorithms},
  \href{https://doi.org/10.1038/s41467-021-27045-6} {Nat. Commun. {\bf 12},
  6961  (2021)}.

\bibitem{Garnerone2010-81}
S. Garnerone, T.~R. de~Oliveira, and P. Zanardi, {\em Typicality in random
  matrix product states}, \href{https://doi.org/10.1103/PhysRevA.81.032336}
  {Phys. Rev. A {\bf 81},  032336  (2010)}.

\bibitem{Garnerone2010-82}
S. Garnerone, T.~R. de~Oliveira, S. Haas, and P. Zanardi, {\em Statistical
  properties of random matrix product states},
  \href{https://doi.org/10.1103/PhysRevA.82.052312} {Phys. Rev. A {\bf 82},
  052312  (2010)}.

\bibitem{Haferkamp2021-2}
J. Haferkamp, C. Bertoni, I. Roth, and J. Eisert, {\em Emergent statistical
  mechanics from properties of disordered random matrix product states},
  \href{https://doi.org/10.1103/PRXQuantum.2.040308} {PRX Quantum {\bf 2},
  040308  (2021)}.

\bibitem{Nahum2017-7}
A. Nahum, J. Ruhman, S. Vijay, and J. Haah, {\em Quantum entanglement growth
  under random unitary dynamics},
  \href{https://doi.org/10.1103/PhysRevX.7.031016} {Phys. Rev. X {\bf 7},
  031016  (2017)}.

\bibitem{Nahum2018-8}
A. Nahum, S. Vijay, and J. Haah, {\em Operator spreading in random unitary
  circuits}, \href{https://doi.org/10.1103/PhysRevX.8.021014} {Phys. Rev. X
  {\bf 8},  021014  (2018)}.

\bibitem{Zhou2019-99}
T. Zhou and A. Nahum, {\em Emergent statistical mechanics of entanglement in
  random unitary circuits}, \href{https://doi.org/10.1103/PhysRevB.99.174205}
  {Phys. Rev. B {\bf 99},  174205  (2019)}.

\bibitem{Potter2022-211}
A.~C. Potter and R. Vasseur,  in {\em Entanglement in Spin Chains: From Theory
  to Quantum Technology Applications}, edited by A. Bayat, S. Bose, and H.
  Johannesson (Springer International Publishing, Cham, 2022), pp.\ 211--249.

\bibitem{Fisher2023-14}
M.~P. Fisher, V. Khemani, A. Nahum, and S. Vijay, {\em Random quantum
  circuits}, \href{https://doi.org/10.1146/annurev-conmatphys-031720-030658}
  {Annu. Rev. Condens. Matter Phys. {\bf 14},  335  (2023)}.

\bibitem{You2021-139}
X. You and X. Wu, {\em Exponentially many local minima in quantum neural
  networks}, \href{https://doi.org/10.48550/arXiv.2110.02479} {Proceedings of
  the 38th International Conference on Machine Learning {\bf 139},  12144
  (2021)}.

\bibitem{Liu2022_06}
J. Liu, Z. Lin, and L. Jiang, {\em Laziness, barren plateau, and noise in
  machine learning}, \href{http://arxiv.org/abs/2206.09313} {arXiv:2206.09313
  (2022)}.

\end{thebibliography}

\end{document}